\documentclass{aa}  

\usepackage{graphicx}
\usepackage{txfonts}
\begin{document} 

\title{MY Camelopardalis, a very massive merger progenitor}

   \author{J. Lorenzo \inst{1}
\and I. Negueruela \inst{1}
\and A.K.F. Val Baker \inst{1}\fnmsep\thanks{Now at Department of Physics, University of Malaya, 50603 Kuala Lumpur, Malaysia.}
\and M. Garc\'{\i}a \inst{2} 
\and S. Sim\'on-D\'{\i}az \inst{3,4}
\and P. Pastor \inst{5} 
\and M. M\'endez Majuelos \inst{6}}

   \institute{Departamento de F\'{\i}sica, Ingenier\'{\i}a de Sistemas y 
Teor\'{\i}a de la Se\~nal, Universidad de Alicante, Apdo. 99, E03080 Alicante, Spain
   \and
Departamento de Astrof\'{\i}sica, Centro de Astrobiolog\'{\i}a (CSIC--INTA), Ctra. de Torrej\'{o}n
a Ajalvir, km 4, E28850 Torrej\'{o}n de Ardoz, Madrid, Spain
   \and
Instituto de Astrof\'{\i}sica de Canarias,  V\'ia L\'actea s/n, E38200, La Laguna, Tenerife, Spain 
\and
Departamento de Astrof\'{\i}sica, Universidad de La Laguna, Facultad de F\'isica y Matem\'aticas, Universidad de La Laguna, Avda. Astrof\'isico Francisco S\'anchez, s/n, E38205, La Laguna, Tenerife, Spain
\and
Departamento de Lenguajes y Sistemas Inform\'{a}ticos, Universidad de Alicante, Apdo. 99, E03080 Alicante, Spain 
\and 
Departamento de Ciencias, IES Arroyo Hondo, c/ Maestro Manuel Casal 2, E11520, Rota, C\'adiz, Spain }
   \date{Received ; accepted }

% \abstract{}{}{}{}{} 
% 5 {} token are mandatory
 
  \abstract
   {The early-type binary MY~Cam belongs to the young open cluster Alicante~1, embedded in Cam OB3.}
 {MY~Cam consists of two early-O type main-sequence stars and shows a photometric modulation suggesting an orbital period slightly above one day. We intend to confirm this orbital period and derive orbital and stellar parameters.}
{Timing analysis of a  very exhaustive (4607 points) light curve indicates a period of $1.1754514\pm0.0000015$~d. High-resolution spectra and the cross-correlation technique implemented in the {\sc todcor} program were used to derive radial velocities and obtain the corresponding radial velocity curves for MY~Cam. Modelling with the stellar atmosphere code {\sc fastwind} was used to obtain stellar parameters and create templates for cross-correlation. Stellar and orbital parameters were derived using the Wilson-Devinney code, such that a complete solution to the binary system could be described.}
   {The determined masses of the primary and secondary stars in MY~Cam are $37.7\pm1.6$ and $31.6\pm1.4\:M_{\sun}$, respectively. The corresponding temperatures, derived from the model atmosphere fit, are 42\,000 and 39\,000~K, with the more massive component being hotter. Both stars are overfilling their Roche lobes, sharing a common envelope. }
   {MY~Cam contains the most massive dwarf O-type stars  found so far in an eclipsing binary. Both components are still on the main sequence, and probably not far from the zero-age main sequence. The system is a likely merger progenitor, owing to its very short period.}


   \maketitle
%
%_______________________________________________________________

\section{Introduction}
Numerous studies show the importance of massive stars for the current properties of galaxies \citep[e.g.][]{massey}. Among O-type stars, relatively unevolved massive stars, there is a very high fraction of binaries, where the initial orbital period determines their evolution and final fate. \citet{sana2012} note that 60\% of O-type binaries have a
period shorter than ten days. In close to 90\% of O-type binaries, at least one of the components will fill its Roche lobe during their lifetimes. This leads to interaction with its companion, where mass transfer, followed by common
envelope evolution, may lead to a merger. Some estimates put 25\% as the fraction of binaries that will merge at some point \citep{lang2012}. In particular, when a binary system has an initial orbital period, $P_{{\rm orb}}$,
shorter than two days and mass ratio $q$ lower than 0.6, evolutionary models predict that both components will merge while on the main sequence. This is fate expected for half of the binary systems that evolve, according to Case A, i.e. those in which mass transfer starts during core hydrogen burning \citep{well2001, lang2012}. For smaller $P_{\rm orb}$, 
systems with higher $q$ will also come into contact during H core burning \citep{well2001}.

\citet{sybe1985}  notes that short-period binary systems with $q>0.8$ 
form stable overcontact systems. These systems would be classified as Case A interacting binaries with slow evolution to contact \citep{eggl2000}. Other theoretical studies have also shown that these types of binaries tend to become overcontact systems evolving through slow Case A \citep[e.g.][]{qian2007}.

Despite all these theoretical predictions, our observational constraints are limited. There are only two known binary systems with two main-sequence O-star components presenting $P_{{\rm orb}}<2$~d whose orbits have been solved: V382~Cyg \citep{harr97} and TU~Mus \citep{terr03}. 
Moreover, because the merger phase is so short compared to stellar lifetimes, only two systems have been proposed to have experienced an observed merger event, and neither of them is a massive system: 
V838~Mon could represent an intermediate-mass merger \citep{muna2002, tyle2011b} and 
V1309~Sco may have been a low-mass merger \citep{maso2010, tyle2011a}.

Mergers are the consequence of angular momentum loss in tidally 
synchronised systems with short initial orbital periods \citep{andr2006}. The physics of mergers is poorly understood, but it is believed that they follow a common envelope phase. This phase, when both stars share the same envelope,  is extremely important in close binary evolution, but is still not fully understood \citep{ivan2013}. The results of mergers are expected to have unusual properties, such as very high rotational velocities or unusual surface abundances \citep{lang2012}. 

In recent years, the mergers of high-mass binaries have been proposed as an 
effective mechanism to form very massive stars. Using population synthesis models, \citet{mink2014} estimate that $8^{+9}_{-4}$\% of apparently single early-type stars are the products of a merger. A binary merger origin has been proposed as an explanation for the extremely fast rotation ($v_{{\rm rot}}> 500\:{\rm km}\,{\rm s}^{-1}$) of the O-type star VFTS~102 \citep{jian2013}, though its peculiar radial velocity is strongly suggestive of other types of binary interaction \citep{duft2011}. Dynamically 
induced mergers of very massive binaries in the cores of dense clusters have been proposed 
as an explanation for the formation of very massive stars ($M_{*}>150\,M_{\sun}$) in the core of R136 
\citep{bane2012}. Some of the energetic events known as supernova impostors could also be 
due to the merger of a massive binary \citep{soke2013}. Somewhat less massive mergers have been proposed as the progenitors of peculiar O-type stars with high magnetic fields \citep[see references in][]{lang2012} or very massive runaway stars \citep{vanb2009}.

MY~Cam (BD +56$^{\circ}$864 = GSC~3725-0498) is the brightest star in the open cluster Alicante~1, a sparsely populated, very young open cluster embedded in 
the Cam OB3 association  \citep{negu2008}. This association has a distance estimate of $\approx 4$~kpc, 
which places it in the Cygnus arm. The Galactic coordinates of MY~Cam are
$l=146\fdg27$, $b=+3\fdg14$. 
MY~Cam was classified as O6nn by \citet{morg1955}. Higher quality spectroscopic observations later
showed that it was a double-lined spectroscopic binary \citep{negu2003}. The two components seemed to have similar spectral types, close to O6\,V, while the large separation in radial velocity 
suggested a very short orbital period \citep{negu2003}. Using data from the Northern Sky 
Variability Survey \citep{wozn2004}, \citet{grea2004} noted a periodicity of $\approx1.18$~d 
in the lightcurve and suggested it was due to elliptical  variations. If this periodicity is the 
orbital period of the binary, then MY~Cam would represent an excellent candidate for a very massive merger progenitor. In this paper, 
we derive the properties of the two components of MY~Cam and find that it fulfils all criteria to be 
considered as a very massive merger progenitor: it is a  high-mass binary system with two early-O type 
components on the main sequence, a very short period, very high rotational
velocity due to synchronisation and components already filling their Roche lobes.

%__________________________________________________________________

\section{Observations}

Spectroscopic observations of MY~Cam were obtained between 23 and 31 December 2004, using {\sc foces} 
(Fibre Optics Cassegrain \'Echelle Spectrograph) mounted on the 2.2-m telescope of the Calar Alto Observatory (Spain). 
The observations were performed in service mode, and sky conditions were very variable. The log of 
observations is displayed in Table~\ref{log}, where we have assigned a number
to each spectrum, sorted by date in ascending order. Heliocentric corrections, calculated using the {\sc rv}
program included in the {\it Starlink} suite, are also listed in Table~\ref{log}.
A total of 63 spectra were obtained, with fixed exposure times of 1800~s. The signal-to-noise ratio 
(S/N) varies strongly between spectra, from $\sim20$ to $\sim50$ per pixel. The spectrograph
covers a wide spectral range between 3780 and 10864\AA. The resolution is slightly variable with wavelength, 
but always $\approx40\,000$ for a 2-pixel resolution element. Given the faintness of the source and the 
spectral response, data for wavelengths shorter than $\sim4000$\AA\ are not useful.

 We reduced all spectra with two methods. Firstly, we used the IDL standard pipeline provided by the observatory. The extracted spectra showed a good blaze correction and order merging, but very poor S/N. The spectra were therefore processed (bias-subtracted, flat-fielded and wavelength-calibrated) using standard {\sc iraf} tools for the reduction of \'echelle spectroscopy. The 99 \'echelle orders were 
blaze-corrected, combined and continuum-normalised to obtain a single spectrum using the pipeline-reduced spectra as a guide.

In addition, on 22 November 2002, we obtained one intermediate-resolution spectrum with the blue arm of the Intermediate dispersion Spectrograph and Imaging System (ISIS) double-beam spectrograph, mounted on the 4.2 m William Herschel 
Telescope (WHT) in La Palma (Spain). The instrument was fitted with the R1200B grating and 
the EEV12 CCD. This configuration covers a spectral range from $\sim4\,000$ to 4\,700\AA\ in the unvignetted 
section of the CCD with a nominal dispersion of 0.25\AA/pixel. The CCD was unbinned in the 
spectral direction and a $1\farcs0$ slit, was used. With this slit the resolution element is 
expected to be smaller than four pixels. The resolving power of our spectra is thus $R\approx5\,000$.

\begin{figure}
   \centering
   \resizebox{\columnwidth}{!}{\includegraphics[clip]{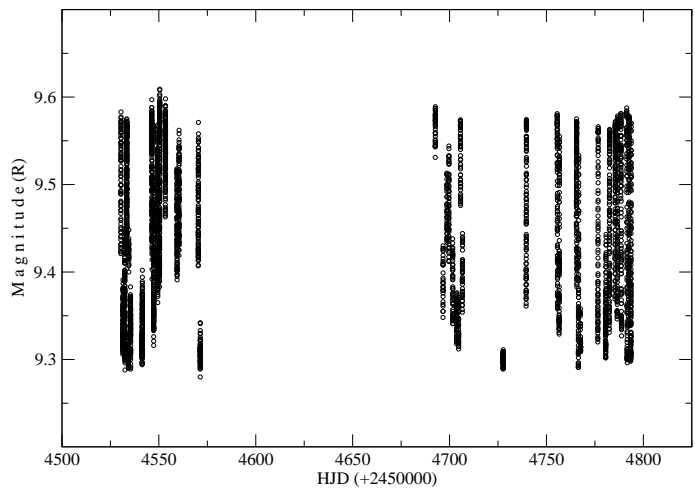}}
      \caption{Photometric light curve, showing the amplitude of the modulation.\label{obs_fot}}
   \end{figure}

Photometric data (Table 2 avalaible online) were obtained with two 8-inch aperture telescopes, a Meade LX200 and a Vixen VISAC, 
with focal-ratios of f/6.3 and f/9, respectively. Observations were made through a Johnson $R$-filter 
with uniform 120-s exposures. We reduced the data with the standard commercial software packages AIP4Win
and Mira Pro. We followed the standard procedures for bias and flat correction. A total of 4607 photometric points were registered, randomly distributed in nights between 4 March and 13 September 2008. 
Photometric values were derived differentially with respect to three reference stars with similar colours 
in the same frame: TYC~3725-797-1,  TYC~3725-00486-1, and TYC~3725-00437-1. The latter is an B1.5\,V star likely 
associated with the open cluster Alicante~1, to which MY~Cam belongs. The photometric light-curve is displayed 
in Fig.~\ref{obs_fot}. A photometric variability of approximately 0.3~mag is apparent.

\begin{table}
\caption{Log of spectroscopic observations sorted by date. \label{log}}      
\centering 
\scalebox{0.94}{   
\begin{tabular}{l c c c}        
\hline\hline           
Number & Date & HJD & Hel. corr. \\   
       &(Day--time)&&(km\:s$^{-1}$) \\
\noalign{\smallskip}
\hline
\noalign{\smallskip}
     1	 & 23--17:47:16 & 2453363.2393  & $-$9.02\\
     2	 & 23--18:21:06 & 2453363.2628  & $-$9.04\\
     3	 & 23--18:53:12 & 2453363.2851  & $-$9.06\\
     4	 & 23--19:25:46 & 2453363.3077  & $-$9.09\\
     5	 & 23--19:57:53 & 2453363.3300  & $-$9.12\\
     6	 & 23--21:07:56 & 2453363.3787  & $-$9.19\\
     7	 & 23--21:40:04 & 2453363.4010  & $-$9.22\\
     8	 & 23--22:12:26 & 2453363.4234  & $-$9.26\\
     9	 & 23--22:44:34 & 2453363.4458  & $-$9.29\\
    10	 & 23--23:59:35 & 2453363.4978  & $-$9.39\\
    11	 & 23--00:31:42 & 2453363.5201  & $-$9.42\\
    12	 & 23--01:04:04 & 2453363.5426  & $-$9.45\\
    13	 & 23--01:36:12 & 2453363.5649  & $-$9.47\\
    14	 & 23--02:10:24 & 2453363.5887  & $-$9.49\\
    15	 & 23--02:42:30 & 2453363.6110  & $-$9.51\\
    16	 & 23--03:15:39 & 2453363.6340  & $-$9.53\\
    17	 & 23--03:47:46 & 2453363.6563  & $-$9.55\\
    18	 & 23--04:20:08 & 2453363.6788  & $-$9.56\\
    19	 & 23--04:53:01 & 2453363.7016  & $-$9.56\\
    20	 & 23--05:25:23 & 2453363.7241  & $-$9.56\\
    21	 & 24--17:43:32 & 2453364.2366  & $-$9.39\\
    22	 & 24--18:15:38 & 2453364.2589  & $-$9.41\\
    23	 & 24--18:48:36 & 2453364.2818  & $-$9.44\\
    24	 & 24--19:20:42 & 2453364.3041  & $-$9.46\\
    25	 & 24--22:10:18 & 2453364.4219  & $-$9.62\\
    26	 & 24--22:58:14 & 2453364.4552  & $-$9.72\\
    27	 & 24--23:30:20 & 2453364.4774  & $-$9.75\\
    28	 & 24--00:02:27 & 2453364.4997  & $-$9.79\\
    29	 & 24--01:07:51 & 2453364.5452  & $-$9.84\\
    30	 & 24--01:40:01 & 2453364.5675  & $-$9.87\\
    31	 & 24--02:12:07 & 2453364.5898  & $-$9.89\\
    32	 & 24--02:44:59 & 2453364.6126  & $-$9.93\\
    33	 & 24--03:17:29 & 2453364.6352  & $-$9.94\\
    34	 & 24--03:49:37 & 2453364.6575  & $-$9.95\\
    35	 & 24--04:21:44 & 2453364.6798  & $-$9.95\\
    36	 & 24--04:54:36 & 2453364.7026  & $-$9.96\\
    37	 & 24--05:26:44 & 2453364.7249  & $-$9.95\\
    38	 & 30--18:54:53 & 2453370.2857  & $-$11.78\\
    39	 & 30--19:27:01 & 2453370.3080  & $-$11.81\\
    40	 & 30--20:02:23 & 2453370.3325  & $-$11.84\\
    41	 & 30--22:55:48 & 2453370.4529  & $-$12.05\\
    42	 & 30--23:27:55 & 2453370.4752  & $-$12.08\\
    43	 & 30--01:57:59 & 2453370.5794  & $-$12.18\\
    44	 & 30--02:30:05 & 2453370.6017  & $-$12.22\\
    45	 & 30--03:02:13 & 2453370.6240  & $-$12.23\\
    46	 & 30--03:34:59 & 2453370.6468  & $-$12.24\\
    47	 & 30--04:08:01 & 2453370.6697  & $-$12.25\\
    48	 & 30--04:40:08 & 2453370.6920  & $-$12.25\\
    49	 & 30--05:14:30 & 2453370.7159  & $-$12.25\\
    50	 & 31--18:24:00 & 2453371.2641  & $-$12.13\\
    51	 & 31--18:58:16 & 2453371.2879  & $-$12.16\\
    52	 & 31--19:33:25 & 2453371.3123  & $-$12.19\\
    53	 & 31--20:06:19 & 2453371.3352  & $-$12.22\\
    54	 & 31--21:01:01 & 2453371.3732  & $-$12.29\\
    55	 & 31--21:33:08 & 2453371.3955  & $-$12.32\\
    56	 & 31--00:33:27 & 2453371.5207  & $-$12.51\\
    57	 & 31--01:05:35 & 2453371.5430  & $-$12.53\\
    58	 & 31--02:09:49 & 2453371.5876  & $-$12.57\\
    59	 & 31--02:42:54 & 2453371.6105  & $-$12.59\\
    60	 & 31--03:15:01 & 2453371.6329  & $-$12.61\\
    61	 & 31--03:47:08 & 2453371.6552  & $-$12.62\\
    62	 & 31--04:19:43 & 2453371.6778  & $-$12.62\\
    63	 & 31--04:51:54 & 2453371.7001  & $-$12.62\\
\noalign{\smallskip}
\hline                        
\end{tabular}}
\end{table}

%______________________________________________________________

\section{Spectroscopic analysis: radial velocity determination}

\begin{figure*}
   \centering
\resizebox{\textwidth}{!}{\includegraphics[clip, angle=+90]{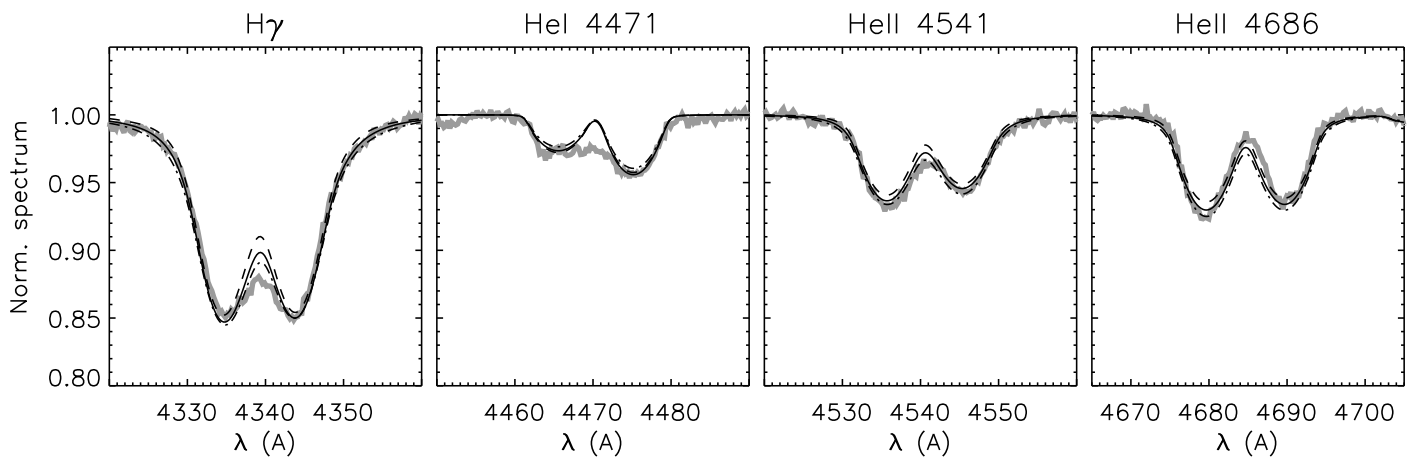}}
      \caption{Some representative lines in the spectrum of MY~Cam and some model line fits. Solid line:  $T_{\rm eff 1}$$=42\,000$~K, $\log\,g_{\rm 1}$ $= 3.9$;
$T_{\rm eff 2}$ $= 39\,000$~K, $\log\,g_{\rm 2}$ $= 4.1$.  Dashed line: $T_{\rm eff 1}$$=40\,000$~K, $\log\,g_{\rm 1}$ $= 3.7$;
$T_{\rm eff 2}$ $= 38\,000$~K, $\log\,g_{\rm 2}$ $= 3.9$. Dashed-dotted line:  $T_{\rm eff 1}$$=44\,000$~K, $\log\,g_{\rm 1}$ $= 4.1$;
$T_{\rm eff 2}$ $= 41\,000$~K, $\log\,g_{\rm 2}$ $= 4.3$. \label{model}}
   \end{figure*}

The spectrum of MY~Cam shows strong He\,{\sc ii}~4686\AA\ absorption, a
characteristic of O-type main-sequence stars. The He\,{\sc ii}~4542 and 5411\AA\ 
lines are also strongly in absorption. A very weak N\,{\sc iii}~4634-40-42\AA\ emission 
complex may be guessed in the spectra with better S/N. According to the classical criteria 
for O-type spectral classification
by \cite{walb1990}, the ratio between the He\,{\sc ii}~4542\AA\ and He\,{\sc i}~4471\AA\ lines 
is larger than unity for spectral types earlier than O7. Both stars show this characteristic, 
with the ratio being larger for one of the stars that we will from now on identify as the primary. Because of blending and the possibility of the Struve-Sahade effect being present (cf. \citealt{lind2007}), 
accurate spectral types cannot be given. However, we can confidently claim that the primary is at 
least as early as O5.5\,V and the secondary is earlier than O7\,V (see Fig.~\ref{model}).

With the purpose of applying the technique of cross-correlation for the determination of radial velocities, 
we have synthesised two templates
using the {\sc fastwind} model atmosphere code \citep{puls05,sant97}. The WHT spectrum, which has a much 
better S/N than any of the {\sc foces} spectra, was used to derive the best-fit parameters. The separation 
between the two components is sufficient to allow for two synthetic spectra to be fitted for O-type stars, the effective temperature is determined from the ionisation balance of \ion{He}{i} and \ion{He}{ii}. There is a small degree of degeneracy between values of $T_{{\rm eff}}$ and $\log \,g$. We therefore used the grid of  {\sc fastwind} models of solar metallicity presented in \citet{sim2011}, and obtained fits by eye. The best-fit model and two models that define the error bars are shown in Fig.~\ref{model}. From this, we conclude that the best fitting parameters are:

$T_{\rm eff 1} =42\,000\pm1\,500$~K, $\log\,g_{\rm 1}$ $= 3.90\pm0.15$

$T_{\rm eff 2} = 39\,000\pm1\,500$~K, $\log\,g_{\rm 2}$ $= 4.10\pm0.15$\, ,

where the error bars indicate the range of parameters that provide acceptable fits  (hereafter, we will use the subindex 1 for the primary star and the subindex 2 for the secondary star). These parameters are in good agreement 
with the derived spectral types. The rotational velocities were estimated simultaneously with the fitting, finding $v_{\rm rot 1,2} \sin\, i$ $= 280\pm20\:{\rm km}\,{\rm s}^{-1}$. 

On the observational scale of \citet{mart2005}, the temperatures 
correspond to spectral types O4.5\,V and O6\,V. We must note that the stars on the scale of \citet{mart2005} are assumed to have $\log\,g$ $= 3.9$. Our values of  $\log\,g$ have not been corrected for the effect of the centrifugal force, which, at the rotational velocities observed, should be by about 0.05~dex in the direction of higher gravities. The model fit is better when the He abundance of the more massive component is increased from the usual $Y_{{\rm He}}=0.10$ to $Y_{{\rm He}}=0.15$. In any case, we must stress that {\sc fastwind} assumes spherical symmetry, an approximation that, as we will see below, is not well justified in such a close binary.

%\begin{figure}
%   \centering
%\resizebox{\columnwidth}{!}{\includegraphics[angle=+90,clip]{templates_my.eps}}
%      \caption{ \label{templates}}
%   \end{figure}

To derive the radial velocities for both components of the binary system, we 
used the technique of cross-correlation in two dimensions developed by \cite{zuck1994} 
and implemented in the {\sc todcor} program. We cross-correlated our observational spectra 
against the two previously synthesised templates. The spectral range of 4400--5585\AA\ was chosen for the analysis, since it includes the main He\,{\sc ii} lines (4542, 4686 and 5411\,\AA)
as well as H$\beta$. These are the only lines clearly detectable in all the {\sc foces} spectra, as all \ion{He}{i} lines become too shallow as soon as the system moves out of eclipse. 
We rebinned all the spectra and the synthetic templates to 9000 bins. The apodization 
factor used in the process was 0.3, meaning that the spectra are smoothed with a cosine bell 
over the 15\% of their length closer to the edges. All radial velocities measured
were corrected to the heliocentric velocity system.

The determination of the uncertainties in the radial velocities is based on
the width of the peak in the cross-correlation function. In the case of MY~Cam, 
because of the very large rotational velocity, the lines are very wide, and therefore uncertainties around 100~km\:s$^{-1}$ are expected. Given the 
size of the bin (around 8~km\:s$^{-1}$), the uncertainties could be reduced
using a larger number of lines for the two-dimensional correlation algorithm. 
Unfortunately, because of the poor S/N of many of the spectra, we could not work with more lines. 
In many spectra, for wavelengths shorter 
than 4300\AA, lines and continuum are practically indistinguishable. As mentioned previously, the He\,{\sc i} lines are only visible when both stars are 
found at the systemic velocity
and the lines of both components are super-imposed. The only other line that is clearly seen is 
H$\alpha$, but we preferred not to include it in the analysis
because, even for stars on the main sequence, it can be significantly affected by stellar wind effects.

\begin{figure}
   \centering
   \resizebox{\columnwidth}{!}{\includegraphics[clip]{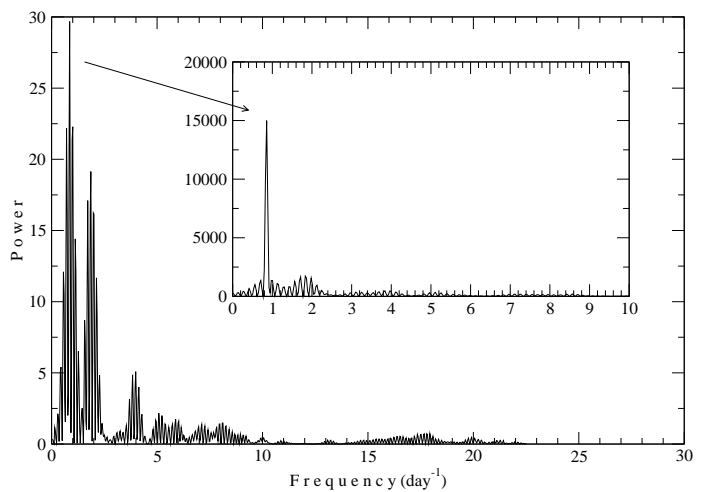}}
      \caption{Results of the Lomb-Scargle periodogram for the radial velocities. The inset shows the result of applying the {\sc clean}
algorithm. \label{periodogram}}
   \end{figure}

In spite of the paucity of spectral features used for the analysis, the results are reliable. 
We used the {\sc period} program inside the {\it Starlink} suite to perform a search for periodicities 
in the radial velocities. The Lomb-Scargle algorithm \citep{lomb1976,scar1982} gives a spectroscopic 
period of $1.17\pm0.04$~d, which is in excellent agreement with the photometric period. This value is 
confirmed with the {\sc clean} algorithm \citep{robe1987}, which removes periods caused by the 
window function (see Fig.~\ref{periodogram}).

\section{Combined analysis: photometry and spectroscopy}

\begin{figure}
   \centering
   \resizebox{\columnwidth}{!}{\includegraphics[clip]{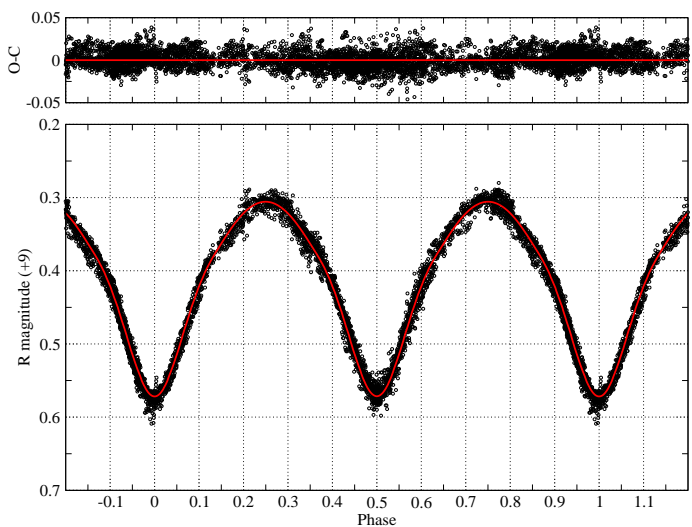}}
      \caption{Light curve model fitted to the observational data. Residuals
are displayed in the top panel. \label{lc_res}}
   \end{figure}

A combined analysis of the photometric data and the radial velocities was made using the Wilson-Devinney code  \citep{wils1971,wils1976,wils1990} in its 2010 version. 
Because of the low signal-to-noise ratio of the spectra and the very high quality of the photometric data (which cover 
the whole orbital period with excellent sampling), we chose to analyse the radial velocity curve and light curve separately, 
following the method described in \citet{ribas2005}. The radial velocity curve and light curve are analysed with an iterative 
procedure until convergence is reached for all free parameters. The criterion for convergence adopted is that for three consecutive 
iterations all adjustable parameters must be within two standard deviations. Once convergence is reached, five solutions are derived 
by varying the parameters within the standard deviation and fitting the observations again. We choose the fit with the smallest dispersion as a final solution. The 
initial parameters were derived from the spectroscopic analysis in the previous section and the stellar parameters 
obtained from the atmospheric model. The first steps of the procedure
rely on finding the period and the zero time of ephemeris, using the photometric data. From the folded light curve, we classify MY~Cam as 
a contact binary with a short period (around a day), so we can
assume that the system is circularised and synchronised. The eccentricity can then be assumed to be zero. The location in orbital phase of the primary and secondary eclipses will be exactly 
0.0 and 0.5, respectively. At phase 0.0, the primary component is being eclipsed by the secondary component, 
while at phase 0.5, the primary component eclipses the secondary component. This is barely perceivable 
in the light curve (Fig.~\ref{lc_res}), where 
the minimum at phase 0.0 is very slightly deeper than the minimum at phase 0.5.To quantify this difference, we average all photometric data points between $\phi=0.99$ and~$0.01$ (average $R=9.5743$), 
and all points between $\phi=0.49$ and~0.51 (average $R=9.5683$). The difference between the two minima is thus 0.0060~mag.

The implementation of the Wilson-Devinney code requires some previous assumptions. Given the stellar parameters
found from the {\sc fastwind} analysis and the extremely short orbital period, we assume that MY~Cam must be a
contact system and therefore share a common equipotential surface. 
The effective temperatures of both components were fixed to the value found from the {\sc fastwind} 
analysis for the hottest component, $T_{{\rm eff}}=42\,000$~K. Every component 
is considered to be divided in a $40 \times 40$ grid of surface elements. A reflection model is included in the code \citep{wils1990}
as well. A square root limb-darkening law 
is applied, as it is more precise than the linear law \citep{hamm1993} and is the most adequate approximation for radiative stars \citep{diaz1995}.
Gravity darkening and ellipsoidal effects are evaluated because the components must be very close to each other.
Finally, the bolometric albedos are taken to be equal to unity, considering that the stellar atmospheres are in 
radiative equilibrium
\citep{vonz1924}. Other contraints applied are described in mode 1 of the Wilson-Devinney code.

The quadrature phases in a circular orbit happen at phases 0.25 and 0.75.
At these phases, the stars are not eclipsed, and we receive the maximum flux from the binary system 
(see Fig.~\ref{lc_res}). In the phenomenological classification, the light curve of MY~Cam corresponds 
to an eclipsing variable of the EW type. The curve displayed in Fig.~\ref{lc_res} shows a 
continuous variation of brightness, with minimal difference between the two eclipse depths. The model curve fitted to the 
observational data is also shown in Fig.~\ref{lc_res}. The standard deviation of the fit is only 0.011~mag, 
and the residuals are always smaller than 0.05~mag, demonstrating an excellent fit. 

%At phases 0.1, 0.4, 0.6 and 0.8, we can observe a very small hump in the model light curve, deviating from a sinusoidal shape. This is a  consequence of the distortion from a spherical shape, i.e. the ellipsoidal effect.

From the period and the zero time of ephemeris, we can write the linear ephemeris equation, where the epoch of 
successive times of primary-eclipse minima (phase zero), $T_{{\rm min}}$, is calculated as: 

\begin{multline*}
 T_{\rm min} = {\rm HJD}~(2454518{.}71500\pm0{.}00021)\\
 + (1^{\rm d}{.}1754514\pm0^{\rm d}{.}0000015)\times E \:\,
\end{multline*}

where $E$ is the integer value of the number of orbital cycles.

\begin{figure}
   \centering
    \resizebox{\columnwidth}{!}{\includegraphics[clip]{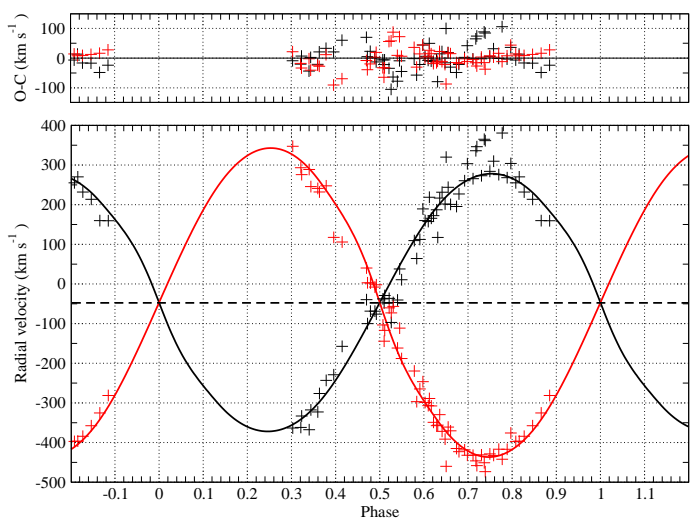}}
      \caption{Radial velocity curves fitted to the observational data and displayed against orbital phase 
(black line: primary star; red line: secondary star) fitted to the observational data.
The dashed line corresponds to the systemic velocity.
The residuals are displayed in the top panel. \label{rvc_res}}
   \end{figure}

The radial velocity points are displayed in Fig.~\ref{rvc_res} together with the final fitted velocity 
curves. The standard deviations are 42\:km\:s$^{-1}$ for the primary star (black curve) and 33\:km\:s$^{-1}$ 
for the secondary star. The residuals, also shown in Fig.~\ref{rvc_res},
reach a maximum value of 100\:km\:s$^{-1}$, in agreement with the uncertainties in
the radial velocities. 
%The Rossiter-McLaughlin effect \citep{ross1924, mcla1924} can just be seen 
%before the eclipses ($\phi$ = 0.4--0.5 and 0.9--1.0), being more apparent in the curve corresponding 
%to the primary star at around phase 0.95, exactly when the primary star is moving away from the observer 
%until it produces the primary eclipse. 
The radial velocities sorted by orbital phase and their residuals 
are listed in Table~\ref{rv_res}. Residuals significantly larger than the standard deviation are found at 
phases around~0.5, owing to the blending of the spectral lines at these phases.

\begin{table}
\setcounter{table}{2}
\caption{Radial velocities and residuals for 
both components of MY~Cam, sorted by orbital phase. \label{rv_res}}      
\centering 
\scalebox{0.92}{   
\begin{tabular}{c c c c c c}        
\hline\hline   
\noalign{\smallskip}        
Number&Phase & RV$_{\rm 1}$ &O-C$_{\rm 1}$& RV$_{2}$&O-C$_{\rm 2}$\\   
   &  &(km\:s$^{-1}$)&(km\:s$^{-1}$) &(km\:s$^{-1}$)&(km\:s$^{-1}$)\\
\noalign{\smallskip}      
\hline
\noalign{\smallskip}      
50 &  0.3029  &  -363  &  -8	&    347 &   21 \\
21 &  0.3211  &  -361  & -20	&    293 &  -17 \\
51 &  0.3231  &  -333  &   6	&    275 &  -33 \\
22 &  0.3401  &  -367  & -43	&    288 &   -2 \\
52 &  0.3439  &  -317  &   2	&    245 &  -40 \\
23 &  0.3596  &  -322  & -20	&    241 &  -23 \\
53 &  0.3634  &  -276  &  20	&    232 &  -27 \\
24 &  0.3786  &  -243  &  33	&    247 &   11 \\
54 &  0.3957  &  -228  &  20	&    117 &  -90 \\
55 &  0.4147  &  -157  &  60	&    105 &  -69 \\
38 &  0.4701  &   -39  &  69	&     40 &   -8 \\
 1 &  0.4722  &  -100  &   5	&      3 &  -38 \\
25 &  0.4789  &   -68  &  22	&      2 &  -19 \\
39 &  0.4890  &   -68  &   1	&     -7 &    2 \\
 2 &  0.4922  &   -76  & -12	&     -2 &   19 \\
26 &  0.5072  &   -37  &  -5	&   -103 &  -32 \\
40 &  0.5099  &   -47  & -21	&   -144 &  -64 \\
 3 &  0.5112  &   -29  &  -5	&   -117 &  -32 \\
56 &  0.5213  &   -36  & -34	&    -60 &   56 \\
27 &  0.5261  &   -97  &-105	&    -73 &   57 \\
 4 &  0.5305  &   -46  & -63	&    -56 &   87 \\
57 &  0.5403  &   -40  & -77	&   -161 &    9 \\
28 &  0.5451  &    37  &  -9	&   -111 &   72 \\
 5 &  0.5494  &    10  & -44	&   -188 &    6 \\
58 &  0.5782  &   109  &  -1	&   -219 &   36 \\
29 &  0.5838  &    64  & -57	&   -296 &  -30 \\
 6 &  0.5909  &   112  & -21	&   -264 &   14 \\
59 &  0.5977  &   189  &  43	&   -246 &   44 \\
30 &  0.6028  &   159  &   6	&   -299 &    0 \\
 7 &  0.6099  &   157  &  -7	&   -306 &    4 \\
41 &  0.6124  &   218  &  49	&   -288 &   26 \\
60 &  0.6168  &   165  & -10	&   -307 &   14 \\
31 &  0.6218  &   172  & -10	&   -348 &  -18 \\
 8 &  0.6289  &   187  &  -6	&   -357 &  -15 \\
42 &  0.6314  &   117  & -79	&   -355 &  -10 \\
61 &  0.6358  &   216  &  14	&   -329 &   22 \\
32 &  0.6412  &   231  &  22	&   -371 &  -11 \\
 9 &  0.6480  &   201  & -15	&   -391 &  -21 \\
43 &  0.6504  &   320  &  99	&   -460 &  -87 \\
62 &  0.6550  &   243  &  18	&   -360 &   17 \\
33 &  0.6604  &   200  & -30	&   -370 &   14 \\
63 &  0.6740  &   195  & -48	&   -415 &  -14 \\
34 &  0.6794  &   227  & -21	&   -423 &  -17 \\
10 &  0.6923  &   260  &   2	&   -418 &   -2 \\
35 &  0.6984  &   303  &  41	&   -431 &  -11 \\
11 &  0.7113  &   265  &  -2	&   -431 &   -4 \\
36 &  0.7178  &   336  &  64	&   -446 &  -15 \\
44 &  0.7201  &   346  &  74	&   -458 &  -26 \\
12 &  0.7304  &   272  &  -2	&   -431 &    3 \\
37 &  0.7368  &   365  &  88	&   -451 &  -15 \\
45 &  0.7390  &   361  &  84	&   -473 &  -37 \\
13 &  0.7494  &   284  &   5	&   -428 &    7 \\
46 &  0.7580  &   309  &  31	&   -432 &    3 \\
14 &  0.7696  &   275  &  -0	&   -416 &   16 \\
47 &  0.7774  &   380  & 105	&   -442 &  -11 \\
15 &  0.7886  &   268  &  -3	&   -416 &    7 \\
48 &  0.7969  &   303  &  36	&   -375 &   43 \\
16 &  0.8082  &   255  &  -6	&   -397 &   13 \\
49 &  0.8159  &   270  &  13	&   -396 &    7 \\
17 &  0.8272  &   231  & -16	&   -384 &    9 \\
18 &  0.8463  &   213  & -16	&   -357 &   12 \\
19 &  0.8657  &   159  & -48	&   -325 &   16 \\
20 &  0.8849  &   159  & -23	&   -281 &   27 \\
\noalign{\smallskip}      
 \hline                        
\end{tabular}}
\end{table}

Some radial velocities for the primary star around the second quadrature ($\phi \sim 0.75$) reach 
values clearly above the fitted curve.
These velocities correspond to the spectra numbered as 36, 37 (December 24) and 44, 45, and 47 
(December 30; see Table~\ref{log}). We assume 
that these deviant velocities can be due to the effects of the stellar wind, being more 
noticeable when the component is not eclipsed by the companion star.

%______________________________________________________________
\section{Results}

The combined analysis of the radial velocity curve and photometric light curve 
permits the determination of the stellar and orbital parameters
for the MY~Cam binary system. All the parameters are listed in Table~\ref{par}. It is worth mentioning that the spectroscopic period is only 54 seconds shorter than the photometric period. The difference between the two periods is around 0.05\%. This close agreement guarantees the quality of the results derived.

\begin{table}
\caption{Stellar parameters derived from the combined analysis of the 
radial velocity and photometric light curve. \label{par}}      
\centering 
\scalebox{0.8}{   
\begin{tabular}{l c c }        
\hline\hline
\noalign{\smallskip}           
&Primary star& Secondary star\\
\noalign{\smallskip}  
\hline
\noalign{\smallskip}  
Orbital period (day)&\multicolumn{2}{c}{1.1754514$\pm$0.0000015}\\
Zero point of ephemeris (HJD)&\multicolumn{2}{c}{2454518.71500$\pm$0.00021}\\
Eccentricity&\multicolumn{2}{c}{0 (assumed)}\\
Inclination ($^{\circ}$)&\multicolumn{2}{c}{62.59$\pm$0.04}\\
Longitude of periastron ($^{\circ}$)&90&270\\
Systemic velocity (km s$^{-1}$)&\multicolumn{2}{c}{$-47\pm$4}\\
Semi-amplitude of velocity (km s$^{-1}$)&335$\pm$7&400$\pm$8\\
Semimajor axis (R$_{\sun}$)& \multicolumn{2}{c}{19.24$\pm$0.26}\\
Surface normalised potential&\multicolumn{2}{c}{3.4853$\pm$0.0011}\\
Mass ($M_{\sun}$)&37.7$\pm$1.6&31.6$\pm$1.4\\
Mass ratio ($M_{2}/M_{1}$)&\multicolumn{2}{c}{0.839$\pm$0.027}\\
Mean radius ($R_{\sun}$)&7.60$\pm$0.10&7.01$\pm$0.09\\
Polar radius ($R_{\sun}$)&7.13$\pm$0.09&6.57$\pm$0.09\\
Point radius ($R_{\sun}$)&9.77$\pm$0.20&9.08$\pm$0.19\\ 
Side radius ($R_{\sun}$)&7.50$\pm$0.10&6.88$\pm$0.09\\
Back radius ($R_{\sun}$)&8.09$\pm$0.11&7.49$\pm$0.10\\
Projected rotational velocity (km s$^{-1}$)&290$\pm$4&268$\pm$3\\
Surface effective gravity ($\log\,g$) &4.251$\pm$0.022&4.245$\pm$0.022\\
Luminosity ratio ($L_{2}/L_{1}$)&\multicolumn{2}{c}{0.8477$\pm$0.0015}\\
\noalign{\smallskip}  
\hline                        
\end{tabular}}
\end{table}

\begin{figure}[!ht]
\begin{center}
\resizebox{\columnwidth}{!}{\includegraphics[clip]{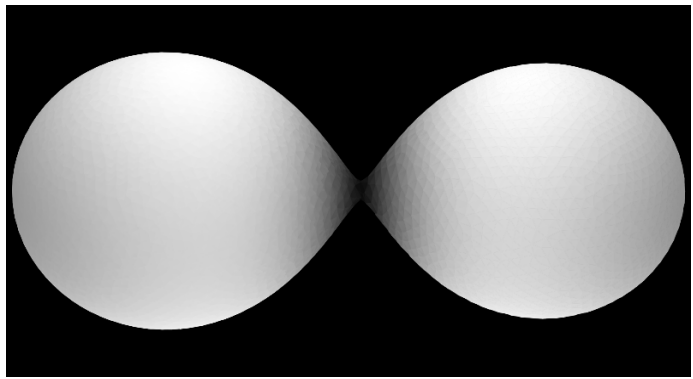}}
\end{center}
\caption{Representative drawing of MY~Cam to scale at quadrature phase, created with the {\it PHOEBE} 2.0-alpha code via the Python interface.}
\label{mycam_quad}
\end{figure}

 Close binaries suffer significant deformations in their shape, which is noticeable in the radii 
 shown in Table~\ref{par}. The point radius is defined as the equatorial radius in the direction 
 to the centre of the other component. The solution provides a value for the point radius, indicating a separation
 between the two stars of $0.39\:R_{\sun}$, but this small distance is compatible with zero within the
 uncertainties derived (see Table~\ref{par}). Therefore, the solution is consistent with the assumption that both stars are overfilling their Roche lobes. 
 Thus, the solution found is compatible with mass transfer already taking place, as suggested by the shape of the lightcurve. A schematic 
 drawing of MY~Cam is displayed in Fig.~\ref{mycam_quad}, showing the shape of both components at quadrature phase.

In addition, since the mass ratio is still significantly different from unity, 
at $q=0.84$, we can conclude that mass transfer has only started recently. The surface gravities are difficult 
to compare, because the {\sc fastwind} model assumes spherical symmetry, while the Wilson-Devinney code takes 
into account that gravity is varying throughout the stellar surface. According to the Wilson-Devinney model, 
the maximum value for gravity is achieved
in the pole of the star ($\log\,g=4.25$), with the minimum value occurring at the point of maximum 
stellar radius ($\log\, g=4.03$). Interestingly, in spite of the very different physics used in the atmosphere 
model and the binary solution, the values obtained for the rotational velocities are essentially the same, 
within their respective errors. 

Examining the subclasses of Case A listed by \citet{eggl2000}, MY~Cam would be
included in subclass AS (slow evolution to contact), because both components have already 
reached contact well before leaving the main sequence. Unfortunately,
evolutionary models presented by \citet{nels2001} do not include binaries 
with massive components and very short periods, such as MY~Cam.  

The derived systemic velocity of $-47\pm4\:{\rm km\phi}\,{\rm s}^{-1}$ is fully consistent with 
the heliocentric velocities of several members
of Cam~OB3 measured by \citet{rubi1965}, which cluster around $-40\:{\rm km}\,{\rm s}^{-1}$. 
This is to be expected for such a massive system in a small-size cluster, where no other stars 
would have a comparable mass necessary for dynamical interactions to provide it with a peculiar velocity.  
The next most massive star in the cluster (and only other high-mass star) is LS~I~$+57\degr$138, with
spectral type O7\,V \citep{negu2008}. 

\section{Discussion and conclusions}

The masses found for the two components of MY~Cam are very high, and, to the best of our knowledge, are the highest masses 
ever derived for main-sequence O-type stars in an eclipsing binary. The literature offers a few examples that can be used 
as comparison systems. For instance, \citet{niem2006} determined the mass of the primary star in V662~Car ($P_{{\rm orb}}=1.41$~d) 
to be around $30\:M_{\sun}$. This star is spectroscopically catalogued as O5.5\,V, which is a cooler (later) star than the primary 
of MY~Cam, and thus less massive. The primary of V1182~Aql ($P_{{\rm orb}}=1.62$~d) \citep{maye2005}
is very similar to the primary in V662~Car, with a mass 
of $31\:M_{\sun}$ and of the same spectral type, O5.5\,V . 

The secondary of MY~Cam can be compared to the primary in V3903~Sgr (O7\,V). This system has an orbital period of 1.74~d, and the mass of the primary is $27\:M_{\sun}$ \citep{vaz1997}. Similarly, the O7\,V 
primary in V382~Cyg ($P_{{\rm orb}}=1.89$~d) has a mass of $28\,M_{\sun}$ \citep{yasa2013}. Again, the secondary of MY~Cam is hotter and heavier than both stars. 

A system with more massive O-type components is DH~Cep, although its orbital period is longer (2.11 days). The masses derived by \citet{hild1996} are around $33\:M_{\sun}$ for the primary and 
$30\:M_{\sun}$ for the secondary. Again, the secondary of MY~Cam is similar in mass to the primary in DH~Cep, while the primary in MY~Cam is even earlier and more massive. We note that DH~Cep is not an eclipsing
binary, and so the masses of its components were determined using other methods of analysis 
\citep{hild1996}.

All the systems listed as comparisons are 
detached eclipsing binaries, while MY~Cam is an early-type overcontact binary. In fact, the 
mean radii of both components in MY~Cam are smaller than the radii derived for stars of similar (or even later) 
spectral types in these 
other binary systems. This difference is very likely due to proximity effects, such as ellipsoidal effects, 
high rotational velocity and tidal effects \citep{zahn1977}. 

Based on the high effective gravities derived for both components, a second possible interpretation to 
the small radii of the two components could be that the stars are very close to the zero-age main sequence (ZAMS). 
The almost complete absence of \ion{N}{iii} emission in their spectra could support this interpretation, 
though the high rotational velocity also contributes to making \ion{N}{iii} emission undetectable. Moreover, 
the strength of \ion{He}{ii}~4686\AA, which is deeper than both  \ion{He}{ii}~4542\AA\ and \ion{He}{i}~4471\AA, 
implies that the two components of MY~Cam are O\,Vz stars. Objects thus classified are generally close to the ZAMS, 
and they are believed to present weak stellar winds \citep{walb2009,sab2014}.  A near-ZAMS nature for MY~Cam would 
be in agreement with the models of \citet{well2001}, who find that stars with period $\la1.2$~d should be in 
contact at the ZAMS for mass ratios $q\ga0.7$.

The effective temperatures found for the components of MY~Cam through model atmosphere fitting are 
hotter than expected for the observed spectral type. \citet{morg1955} found a combined spectral type of O6nn, 
while \citet{negu2008} estimated that one of the components was slightly earlier and the other slightly 
later than O6\,V. The temperatures derived would imply approximate spectral 
types O4\,V and O6\,V, according to the calibration of \citet{mart2005}. Even though the  Struve-Sahade effect (the strengthening of the secondary spectrum
of the binary when it is approaching) may distort the spectral types in massive binaries \citep{gies1997,bagn1999}, 
the high gravity of the stars may also contribute to their higher temperatures. It is also possible that 
reflection effects contribute to increasing the surface temperatures of both components.

Stars more massive than the primary of MY~Cam have been found in binary systems, but these
objects appear as either Of supergiants or WNha stars. The best known examples are the 
eclipsing binaries NGC~3603~A1 ($P_{{\rm orb}} = 3.7724$~d, $M_{1} =116\pm31\,M_{\odot}$, $M_{2} =89\pm16\,M_{\odot}$;
\citealt{schnurr08}) and W20a \citep[$P_{{\rm orb}} = 3.69\pm0.01$~d, $M_{1} =83\pm5\,M_{\odot}$, 
$M_{2} =82\pm5\,M_{\odot}$;][]{bonanos04}, where all the components are WN6ha stars. When such massive 
objects start interacting, the physics of mass transfer is likely to be very complex. Conversely, 
MY~Cam seems to simply be an extremely massive version of slow Case A binary evolution. Even though 
models describing the final fate of such a massive close binary do not exist, analogy with lower-mass 
binaries suggest that common envelope evolution will result in a stellar merger before any of the components 
finish H-core burning. Even if some material is lost from the system during such a process, the product of 
this merger will be a very massive star that will probably display unusual properties \citep[and references 
therein]{ivan2013}.

\begin{acknowledgements}
We thank Francesc Vilardell for many rewarding discussions on the use of the Wilson-Devinney code. 
We also thank the referee, Dirk Terrell, for many insightful comments.
Based on observations collected at the Centro Astron\'omico Hispano 
Alem\'an (CAHA), operated jointly by the Max-Planck Institut für 
Astronomie and the Instituto de Astrof\'{\i}sica de Andaluc\'{\i}a (CSIC). The WHT is operated on the 
island of La Palma by the Isaac Newton Group in the Spanish Observatorio del Roque de Los Muchachos of the Instituto de Astrof\'{\i}sica de Canarias.
We thank Prof.\ Norbert Langer for comments on binary models of merger progenitors.

 This research is partially supported by the Spanish Ministerio de Econo\'{\i}a y Competitividad under 
grant AYA2012-39364-C02-01/02, and the European Union. This research has made use of the Simbad, Vizier 
and Aladin services developed at the Centre de Donn\'ees Astronomiques de Strasbourg, France
\end{acknowledgements}

%-------------------------------------------------------------------

\end{document}